\documentclass[preprint2]{aastex}

\slugcomment{To appear in Astrophys. Journal Supp. Ser.}

\shorttitle{The Morphology of Satellite Galaxies}
\shortauthors{C. M. Gutierrez et al.}

\begin{document}

\title{The Properties of Satellite Galaxies in External Systems.\\ I. Morphology and Structural Parameters}

\author{C. M. Guti\'errez}
\affil{Instituto de Astrof\'{\i}sica de Canarias, E-38205 La Laguna, Tenerife, Spain}
\email{cgc@ll.iac.es}
\author{M. Azzaro}
\affil{Isaac Newton Group of Telescopes, Apdo, 321, E-38700 Santa Cruz de La
Palma, Canary Islands, Spain}
\and
\author{F. Prada}
\affil{Centro Astron\'omico Hispano-Alem\'an, Apdo. 511, E-04080 Almer\'\i
a, Spain}

\begin{abstract}
We present the first results of an ongoing project to study the
morphological, kinematical, dynamical, and chemical properties of
satellite galaxies of external giant spiral galaxies. The sample
of objects has been selected from the catalogue by Zaritsky et al.
(1997). The paper analyzes the morphology and structural parameters of a
subsample of 60 such objects. The satellites span a great variety of
morphologies and surface brightness profiles. About two thirds of the
sample are spirals and irregulars, the remaining third being
early-types. Some cases showing interaction between pairs of
satellites are presented and briefly discussed. 
\end{abstract}


\section{INTRODUCTION}

The standard model for the formation of large structures in the
Universe predicts a hierarchical scenario in which the first generation
of objects created correspond to subgalactic masses (Frenk et al.
1988).  The aggregation of such objects by accretion, merging, etc., over
cosmological timescales would generate the wide variety of structures
observed in the Universe. Mergers between two objects of similar masses
destroy  disks (Barnes \& Hernquist 1991) and seem to be one of the
mechanisms that create ellipticals from spiral galaxies.  A merger between
a giant galaxy and a small satellite tends to heat and thicken the disk
(Vel\'azquez \& White 1999) and  could contribute
to the growth of the bulge and an increase in the S\'ersic $n$
parameter (S\'ersic 1968) moving the galaxy toward earlier types (Walker, Mihos, \&
Herndquist (1996); Aguerri, Balcells, \& Peletier 2001). This accretion of satellites can also
trigger starbursts  (Mihos \& Hernquist 1994), and produce
counterrotating disks (e.g., Thakar \& Ryden 1996). However,
there is some controversy over whether or not the observed structure in
 disk galaxies is compatible with the number of such mergers
expected in the standard model (Toth \& Ostriker (1992); Zaritsky
(1995); Weinberg (1997); Velazquez \& White 1999).

The Milky Way and M31 are examples of bright spiral
galaxies surrounded by several satellites. Semi-analytic models
(Kauffmann, White, \& Guideroni
 1993) and  numerical simulation (Klypin et al.
(1999); Moore et al.  1999) of hierarchical clustering galaxy formation
predict a number of satellites an order of magnitude larger than 
observed in the Local Group. Several alternatives have been proposed to
solve this discrepancy. It has been suggested (Klypin et al. (1999), Blitz et al. 1999)  
that these missing satellites were in the form of the compact high
velocity clouds identified by Braun \& Burton (1999). Bode et al. (2001) have analyzed a
warm dark matter model in which the dark matter particles have non-zero velocities; these tend to suppress the structures on small scales. In
this model  small halos should be formed by fragmentation instead
of by aggregation. Using numerical simulations, Bode et al. show a
reasonable agreement between the predictions of their model and 
observations on the number of satellites, the spatial distribution, and
the epoch of formation. A possible reionization of the Universe at high
redshift could also suppress the formation of small galaxies after that
epoch (Bullock, Kravtsov, \&  Weinberg (2000); Somerville (2001); Benson et al. 2001).

The existence of satellite galaxies in external systems has been known for a long time ($e.\,g.$ Page (1952), Holmberg 1969). The most complete compilation and study of
such objects was conducted by Zaritsky et al. (1997) who presented a
catalogue containing 115 satellites orbiting 69 primary isolated spiral
galaxies. Zaritsky et al. defined satellites as objects at projected
distances $\le 500$ kpc from their primaries, with differences in recessional
velocity $\le 500$ km s$^{-1}$ and  at least 2.2 mag fainter than their
parent galaxy. The number of catalogued satellites per system is very
small (typically 1--2) and for construction of the catalogue they should
correspond to the brightest part of the luminosity function of these
objects (see Pritchet \& van den Bergh 1999). Although the completeness
of this catalogue is hardly understood (see Zaritsky et al. 1993), it
can be estimated to absolute limiting magnitude M$_B\sim -15.5$.
Zaritsky et al., have used the catalogue to demonstrate that the halos
of the primaries extend to at least 200 kpc and have masses larger than
$2\times 10^{12}\ M_\odot$. Other observational studies have been conducted
by Carignan et al. (1997), who have discovered and analyzed eight dwarf
galaxies orbiting the massive lenticular galaxy NGC~5084. This group
shows a nice example of one of the predictions of the standard models:
i.e., there should be a net excess of satellites in retrograde
orbits because, as the numerical simulations show (Quinn \& Goodman 1986), it is more effective
for the parent galaxy to accrete the satellites in direct orbits. Cote
et al. (1997) have discovered 16 and 20 dwarf galaxies in the Sculptor
and the Centaurus group respectively. Some controversy exist about the reality of the 
so-called ``Holmberg effect'', i.e., the fact that satellites tend
to avoid the plane of the primary; while Zaritsky et al., claimed  the
existence of the effect in their sample, Carignan et al., have not found
it in their study of the NGC~5084 group.

In the study of these external satellites it is interesting to compare
their properties with those observed in the Local Group; this will
help us to understand the theoretical models for the formation of
structures in galactic halos. We are conducting an ongoing
observational  program which comprises broad band photometry in the
optical ($B$, $V$, $R$, and $I$) and infrared  ($J$ and $K'$), and in
the H$\alpha$ narrow band for both the parent and the satellite
galaxies, taken from the compilation by Zaritsky et al. (1997). The aim
of this study is to determine the structure,  dynamics, star formation
rates, and stellar populations to constrain the age and metallicity of
such systems.  Here, we concentrate on the morphology and structural
parameters of the satellite galaxies.

\section{THE SAMPLE, OBSERVATIONS, AND DATA REDUCTION}

The observations presented here correspond to a subsample of 60
satellites taken from the compilation by Zaritsky et al. (1997). This
sample comprises most of the objects situated at dec $\ge -22$ degrees.
During several runs between 1997 and 2001 we used the
 0.8 m IAC80$^1$   and the 1 m OGS\footnote{The telescopes are
at the Spanish Teide Observatory on the island of Tenerife and are
operated by the Instituto de Astrof\'\i sica de Canarias} telescopes, 
the 2.5 m Isaac Newton Telescope (INT) and the 4.2 William Herschel
Telescope (WHT) at
the Roque
de Los Muchachos Observatory\footnote{These
telescopes are  operated on the island of La Palma by the Isaac Newton
Group }, and the 2.2 m and 3.5 m telescopes at Calar Alto \footnote{The German-Spanish Astronomical
Centre, Calar Alto is operated by the Max-Planck-Institute for
Astronomy, Heidelberg, jointly with the Spanish National Commission for
Astronomy.}. Typical exposures times were
$\sim$ 30--90 minutes, and $\sim$ 5--10 minutes for the 1 m class and for
larger telescopes respectively. In the analysis presented here, we use
the $R$ filter in most of the cases, although in a few cases we used the
$I$ or the $V$ filter (see below). The typical magnitude limit was $R$ $\sim
24$ mag/arcsec$^2$. The pixel size is different according to the telescope
and instrument used but was typically  in the range 0.3--0.5 arcsec. The
seeing was between 1 and 3 arcsec with a typical value of 2 arcsec.
We performed standard data reduction: (bias subtraction, flat fielding
correction, co-addition of images, sky subtraction, and cleaning of the
remaining cosmic rays).

\section{MORPHOLOGY AND STRUCTURAL PARAMETERS}

The morphological type of the objects was determined by visual inspection, and this was our first criterion for  classification. Using the  ELLIPSE task
in IRAF\footnote{IRAF is the Image Reduction and Analysis Facility,
written and supported by the IRAF programming group at the National
Optical Astronomy Observatories (NOAO) in Tucson, Arizona.} we fitted
ellipses to their isophotes and determined from these the surface
brightness profile. For the bulge we used a S\'ersic law: 

\begin{equation}
 I(r)=I_{\rm e}10^{-b_n[(r/r_{\rm e})^{(1/n)}-1]},
\end{equation} 
where $r$ is the radial coordinate and $b_n=0.868n-0.142$ is defined so
that $r_{\rm e}$ is the radius enclosing half of the light of this
component  (Capaccioli 1989). The de Vaucouleurs profile corresponds to
the particular case $n=4$. For the disk component we used an
exponential law:

\begin{equation} 
I(r)=I_0e^{(-r/r_{\rm d})}.
\end{equation} 

In total there are five free parameters to be determined,  $I_{\rm e}$,
$n$, and $r_{\rm e}$ for the spheroid and $I_0$ and $r_{\rm d}$ for the disk
respectively. However, we do not have  photometric calibration for
our images and so we could not determine the absolute brightness of
each component (characterized by the parameters $I_{\rm e}$ and
$I_0$).

The sizes of most of the galaxies analyzed in this paper are
considerably larger than the seeing (see below). Furthermore, the effect
of  seeing is relevant only in the central parts of the galaxy in
which bulges dominated, the gradients in luminosity usually being higher
and  more dramatic for S\'ersic profiles with high $n$ parameter. In
order to eliminate or greatly reduce the effect, we have excluded from
the fits the  central few arcsec of the profile. The exact size of the
region excluded depends on the size and profile of each galaxy. We have
tested our procedure by changing the region excluded from the fit,
in several cases making a comparison between the parameters derived
with our fits and those obtained with the full treatment of the
seeing (Trujillo et al. 2001), and found that our approach is correct.

Figure 1 shows grayscales image, contour plots, surface brightness
profiles, and the fits of the satellite galaxies analyzed in this
paper. North is up and east is to the left. The field of view depends
on each case and is indicated in the contour plots. In the fits with
both a spheroid and a disk, each component is also shown in the plots.

Table~1 presents the structural and morphological types of the
objects. The columns are 1: the name of the object; 2: $M_B$ the
absolute magnitude in the $B$ band taken from Zaritsky et al. (1997);
3: $D$ the projected distance in kpc between the satellite and the parent
galaxy; 4: the filter used: 5-6: $r_e$ the radius of the bulge in arcsec and kpc
respectively (we have adopted $h_0=0.75$); 7: the $n$ parameter of
the bulge; 8-9: $r_d$ the length scale of the disk in arcsec and kpc
respectively; 10: B/T the bulge to total luminosity ratio computed 
from the structural parameters in columns 5-9, and 11: the
morphological Hubble type.

For the great majority of the objects it was possible to determine
their morphological type. The symbols * and ** quoted in some of the cases
of Table 1 mean uncertain classification and signs of distortion/interaction
respectively.  We wish to make the following remarks on some of the
individual objects:

\begin{itemize}

\item NGC 488c and NGC 2424b are examples of a low surface brightness
galaxies without any structure. Other galaxies in the sample such as
NGC 4030c also have low surface brightness, but in this case there
are knots.

\item There are a few cases of spiral edge-on galaxies in which we were
not able to determine the Hubble type and no satisfactory fit was
found. In these cases we have classified them as S.

\item Many galaxies ($\sim 25$ \%) show signs of distortion, and at
least in two cases interactions (see below).

\item NGC 488b and NGC 2775b are examples of spiral galaxies with very
distorted disks. In the case of NGC 488b only the bulge was fitted. 

\end{itemize}

\section{DISCUSSION}

The results presented here show that satellite galaxies present a
wide variety of morphological types, sizes, and brightness. The Hubble
types go from pure ellipticals to irregulars. About 35
objects have been classified as spiral galaxies.

The largest group in the catalogue is NGC 1961 and its five satellites.
All these satellites are luminous ($-19.6\le M_B \le -18.1$) spiral
galaxies. This group is notably different from the Local Group in which
only M33 and LMC have  magnitude in this range. The other large groups in the
sample are those formed by NGC 4030 and NGC 4541, which have four detected
satellites each. This last is a notable group in which two of the
galaxies show clear interaction (see below), and another, NGC 4541d, shows
strong asymmetries. There are two clear cases of interaction between
pairs of satellites: NGC 2718a and NGC 2718b, and NGC 4541b and
NGC 4541e, respectively. This is illustrated in Figure 2. In the
first case the interaction shows up in the form of a bridge connecting
both galaxies. In the case of the second interacting pair, two tidal
tails seem to emerge from NGC 4541b. One of them points directly to NGC 4541e
and  it seems clear that its origin is the interaction with this
galaxy; the other is nearly in the opposite direction and could reveal
 stripping from a previous orbit. The asymmetries shown by NGC 4541e
could be also related with this interaction. Another notable case is
NGC 4030c classified as Irr, in which several extended objects are
surrounded by a common diffuse structure resembling a case of strong
interaction or merger.

In a forthcoming paper, we will analyze the relationship between the
different structural parameters, and between these parameters and the relative
position and orientation of the satellites with respect to the parent
galaxy. Considering the possible bias, completeness, and sampling of our
data set, we will also conduct a detailed comparison between the
properties found in our systems and those  in the Local Group.

\acknowledgements
We would like to thank to A. Mora, S. Palomares, J. C. Suarez, C. Hoyos,
and M. S. Alonso that   collaborated
with the observations under a summer grant at the IAC.

\section*{FIGURE CAPTIONS}

Fig. 1. Grayscales images, contours, surface brightness profiles and fits for the sample of satellite galaxies analyzed in this paper.

Fig. 2. Interaction between: ($top$) NGC2718a and NGC2718b, and ($bottom$) NGC4541b and NGC4541e.

\newpage
\onecolumn

\begin{table}
\caption{Morphology and structural prameters of satellite galaxies}
\begin{center}
\begin{tabular}{ccccccccccc}
\hline
\multicolumn{1}{c}{Name}  &
\multicolumn{1}{c}{$M_B$}  &
\multicolumn{1}{c}D  &
\multicolumn{1}{c}{filter}  &
\multicolumn{2}{c}{$r_e$}  &
\multicolumn{1}{c}{n}  &
\multicolumn{2}{c}{$r_d$}  &
\multicolumn{1}{c}{B/T}  &
\multicolumn{1}{c}{Hubble} \\
\multicolumn{1}{c}{ }  &
\multicolumn{1}{c}{ }  &
\multicolumn{1}{c}{kpc}  &
\multicolumn{1}{c}{ }  &
\multicolumn{1}{c}{(``)}  &
\multicolumn{1}{c}{kpc}  &
\multicolumn{1}{c}{ }  &
\multicolumn{1}{c}{(``)}  &
\multicolumn{1}{c}{kpc}  &
\multicolumn{1}{c}{ }  &
\multicolumn{1}{c}{type}  \\
\hline
\multicolumn{1}{c} {n259b}  & 
\multicolumn{1}{c} {-17.9}  & 
\multicolumn{1}{c} {376}  & 
\multicolumn{1}{c} {R}  & 
\multicolumn{1}{c} {3.3}  & 
\multicolumn{1}{c} {0.8}  & 
\multicolumn{1}{c} {1.2}  & 
\multicolumn{1}{c} {-}  & 
\multicolumn{1}{c} {-}  & 
\multicolumn{1}{c} {-}  & 
\multicolumn{1}{c}{E}  \\
\hline
\multicolumn{1}{c} {n488a}  & 
\multicolumn{1}{c} {-16.9}  & 
\multicolumn{1}{c} {63}  & 
\multicolumn{1}{c} {R}  & 
\multicolumn{1}{c} {7.6}  & 
\multicolumn{1}{c} {1.0}  & 
\multicolumn{1}{c} {4.3}  & 
\multicolumn{1}{c} {-}  & 
\multicolumn{1}{c} {-}  & 
\multicolumn{1}{c} {-}  & 
\multicolumn{1}{c}{E}  \\
\hline
\multicolumn{1}{c} {n488b}  & 
\multicolumn{1}{c} {-15.5}  & 
\multicolumn{1}{c} {322}  & 
\multicolumn{1}{c} {R}  & 
\multicolumn{1}{c} {12.7}  & 
\multicolumn{1}{c} {1.8}  & 
\multicolumn{1}{c} {1.2}  & 
\multicolumn{1}{c} {-}  & 
\multicolumn{1}{c} {-}  & 
\multicolumn{1}{c} {-}  & 
\multicolumn{1}{c} {S *,**}  \\
\hline
\multicolumn{1}{c} {n488c}  & 
\multicolumn{1}{c} {-15.5}  & 
\multicolumn{1}{c} {116}  & 
\multicolumn{1}{c} {R}  & 
\multicolumn{1}{c} {-}  & 
\multicolumn{1}{c} {-}  & 
\multicolumn{1}{c} {-}  & 
\multicolumn{1}{c} {-}  & 
\multicolumn{1}{c} {-}  & 
\multicolumn{1}{c} {-}  & 
\multicolumn{1}{c} {Irr (LSB)}  \\
\hline
\multicolumn{1}{c} {i1723a}  & 
\multicolumn{1}{c} {-15.7}  & 
\multicolumn{1}{c} {389}  & 
\multicolumn{1}{c} {R}  & 
\multicolumn{1}{c} {-}  & 
\multicolumn{1}{c} {-}  & 
\multicolumn{1}{c} {-}  & 
\multicolumn{1}{c} {-}  & 
\multicolumn{1}{c} {-}  & 
\multicolumn{1}{c} {-}  & 
\multicolumn{1}{c}{Irr}  \\
\hline
\multicolumn{1}{c} {i1723b}  & 
\multicolumn{1}{c} {-17.9}  & 
\multicolumn{1}{c} {135}  & 
\multicolumn{1}{c} {I}  & 
\multicolumn{1}{c} {1.6}  & 
\multicolumn{1}{c} {0.6}  & 
\multicolumn{1}{c} {0.6}  & 
\multicolumn{1}{c} {2.5}  & 
\multicolumn{1}{c} {0.9}  & 
\multicolumn{1}{c} {0.6}  & 
\multicolumn{1}{c}{E/S0}  \\
\hline
\multicolumn{1}{c} {n749a}  & 
\multicolumn{1}{c} {-16.7}  & 
\multicolumn{1}{c} {388}  & 
\multicolumn{1}{c} {R}  & 
\multicolumn{1}{c} {11.5}  & 
\multicolumn{1}{c} {3.3}  & 
\multicolumn{1}{c} {1.7}  & 
\multicolumn{1}{c} {-}  & 
\multicolumn{1}{c} {-}  & 
\multicolumn{1}{c} {-}  & 
\multicolumn{1}{c}{E/S0}  \\
\hline
\multicolumn{1}{c} {n749b}  & 
\multicolumn{1}{c} {-16.1}  & 
\multicolumn{1}{c} {97}  & 
\multicolumn{1}{c} {R}  & 
\multicolumn{1}{c} {3.7}  & 
\multicolumn{1}{c} {1.1}  & 
\multicolumn{1}{c} {0.7}  & 
\multicolumn{1}{c} {-}  & 
\multicolumn{1}{c} {-}  & 
\multicolumn{1}{c} {-}  & 
\multicolumn{1}{c}{E/S0}  \\
\hline
\multicolumn{1}{c} {n772a}  & 
\multicolumn{1}{c} {-18.2}  & 
\multicolumn{1}{c} {32}  & 
\multicolumn{1}{c} {R}  & 
\multicolumn{1}{c} {5.2}  & 
\multicolumn{1}{c} {0.8}  & 
\multicolumn{1}{c} {2}  & 
\multicolumn{1}{c} {-}  & 
\multicolumn{1}{c} {-}  & 
\multicolumn{1}{c} {-}  & 
\multicolumn{1}{c}{E/S0}  \\
\hline
\multicolumn{1}{c} {n772b}  & 
\multicolumn{1}{c} {-15.5}  & 
\multicolumn{1}{c} {390}  & 
\multicolumn{1}{c} {R}  & 
\multicolumn{1}{c} {-}  & 
\multicolumn{1}{c} {-}  & 
\multicolumn{1}{c} {-}  & 
\multicolumn{1}{c} {-}  & 
\multicolumn{1}{c} {-}  & 
\multicolumn{1}{c} {-}  & 
\multicolumn{1}{c}{Sc/Irr}  \\
\hline
\multicolumn{1}{c} {n772c}  & 
\multicolumn{1}{c} {-16.2}  & 
\multicolumn{1}{c} {429}  & 
\multicolumn{1}{c} {R}  & 
\multicolumn{1}{c} {-}  & 
\multicolumn{1}{c} {-}  & 
\multicolumn{1}{c} {-}  & 
\multicolumn{1}{c} {-}  & 
\multicolumn{1}{c} {-}  & 
\multicolumn{1}{c} {-}  & 
\multicolumn{1}{c}{Sa}  \\
\hline
\multicolumn{1}{c} {n895a}  & 
\multicolumn{1}{c} {-14.7}  & 
\multicolumn{1}{c} {19}  & 
\multicolumn{1}{c} {R}  & 
\multicolumn{1}{c} {3.5}  & 
\multicolumn{1}{c} {0.5}  & 
\multicolumn{1}{c} {1.7}  & 
\multicolumn{1}{c} {-}  & 
\multicolumn{1}{c} {-}  & 
\multicolumn{1}{c} {-}  & 
\multicolumn{1}{c}{E (off-centre nuclei) **}  \\
\hline
\multicolumn{1}{c} {n1517a}  & 
\multicolumn{1}{c} {-17.0}  & 
\multicolumn{1}{c} {130}  & 
\multicolumn{1}{c} {R}  & 
\multicolumn{1}{c} {-}  & 
\multicolumn{1}{c} {-}  & 
\multicolumn{1}{c} {-}  & 
\multicolumn{1}{c} {6.8}  & 
\multicolumn{1}{c} {1.6}  & 
\multicolumn{1}{c} {-}  & 
\multicolumn{1}{c}{Sb/Sc}  \\
\hline
\multicolumn{1}{c} {n1620a}  & 
\multicolumn{1}{c} {-17.2}  & 
\multicolumn{1}{c} {227}  & 
\multicolumn{1}{c} {R}  & 
\multicolumn{1}{c} {5.4}  & 
\multicolumn{1}{c} {1.3}  & 
\multicolumn{1}{c} {0.5}  & 
\multicolumn{1}{c} {13.8}  & 
\multicolumn{1}{c} {3.3}  & 
\multicolumn{1}{c} {0.2}  & 
\multicolumn{1}{c}{Sc **}  \\
\hline
\multicolumn{1}{c} {n1620b}  & 
\multicolumn{1}{c} {-17.7}  & 
\multicolumn{1}{c} {203}  & 
\multicolumn{1}{c} {R}  & 
\multicolumn{1}{c} {7.0}  & 
\multicolumn{1}{c} {1.7}  & 
\multicolumn{1}{c} {1.5}  & 
\multicolumn{1}{c} {-}  & 
\multicolumn{1}{c} {-}  & 
\multicolumn{1}{c} {-}  & 
\multicolumn{1}{c}{E/S0}  \\
\hline
\multicolumn{1}{c} {n1640a}  & 
\multicolumn{1}{c} {-16.3}  & 
\multicolumn{1}{c} {402}  & 
\multicolumn{1}{c} {R}  & 
\multicolumn{1}{c} {-}  & 
\multicolumn{1}{c} {-}  & 
\multicolumn{1}{c} {-}  & 
\multicolumn{1}{c} {9.2}  & 
\multicolumn{1}{c} {1.0}  & 
\multicolumn{1}{c} {-}  & 
\multicolumn{1}{c}{Sb/Sc}  \\
\hline
\multicolumn{1}{c} {n1961a}  & 
\multicolumn{1}{c} {-18.8}  & 
\multicolumn{1}{c} {214}  & 
\multicolumn{1}{c} {R}  & 
\multicolumn{1}{c} {1.5}  & 
\multicolumn{1}{c} {0.4}  & 
\multicolumn{1}{c} {0.5}  & 
\multicolumn{1}{c} {11.3}  & 
\multicolumn{1}{c} {2.8}  & 
\multicolumn{1}{c} {0.03}  & 
\multicolumn{1}{c}{Sb/Sc}  \\
\hline
\multicolumn{1}{c} {n1961b}  & 
\multicolumn{1}{c} {-18.1}  & 
\multicolumn{1}{c} {139}  & 
\multicolumn{1}{c} {R}  & 
\multicolumn{1}{c} {-}  & 
\multicolumn{1}{c} {-}  & 
\multicolumn{1}{c} {-}  & 
\multicolumn{1}{c} {4.7}  & 
\multicolumn{1}{c} {1.2}  & 
\multicolumn{1}{c} {-}  & 
\multicolumn{1}{c}{Sb/Sc}  \\
\hline
\multicolumn{1}{c} {n1961c}  & 
\multicolumn{1}{c} {-18.8}  & 
\multicolumn{1}{c} {120}  & 
\multicolumn{1}{c} {R}  & 
\multicolumn{1}{c} {1.0}  & 
\multicolumn{1}{c} {0.3}  & 
\multicolumn{1}{c} {0.4}  & 
\multicolumn{1}{c} {5.9}  & 
\multicolumn{1}{c} {1.6}  & 
\multicolumn{1}{c} {0.1}  & 
\multicolumn{1}{c}{Sb/Sc}  \\
\hline
\multicolumn{1}{c} {n1961d}  & 
\multicolumn{1}{c} {-18.7}  & 
\multicolumn{1}{c} {435}  & 
\multicolumn{1}{c} {R}  & 
\multicolumn{1}{c} {5.0}  & 
\multicolumn{1}{c} {1.4}  & 
\multicolumn{1}{c} {0.6}  & 
\multicolumn{1}{c} {22.8}  & 
\multicolumn{1}{c} {6.3}  & 
\multicolumn{1}{c} {0.3}  & 
\multicolumn{1}{c}{Sb/Sc *}  \\
\hline
\multicolumn{1}{c} {n1961e}  & 
\multicolumn{1}{c} {-19.6}  & 
\multicolumn{1}{c} {319}  & 
\multicolumn{1}{c} {R}  & 
\multicolumn{1}{c} {2.0}  & 
\multicolumn{1}{c} {0.6}  & 
\multicolumn{1}{c} {0.5}  & 
\multicolumn{1}{c} {11.1}  & 
\multicolumn{1}{c} {3.0}  & 
\multicolumn{1}{c} {0.1}  & 
\multicolumn{1}{c}{Sb/Sc}  \\
\hline
\multicolumn{1}{c} {n2424b}  & 
\multicolumn{1}{c} {-15.9}  & 
\multicolumn{1}{c} {182}  & 
\multicolumn{1}{c} {R}  & 
\multicolumn{1}{c} {-}  & 
\multicolumn{1}{c} {-}  & 
\multicolumn{1}{c} {-}  & 
\multicolumn{1}{c} {-}  & 
\multicolumn{1}{c} {-}  & 
\multicolumn{1}{c} {-}  & 
\multicolumn{1}{c}{Irr (LSB)}  \\
\hline
\multicolumn{1}{c} {n2718a}  & 
\multicolumn{1}{c} {-16.5}  & 
\multicolumn{1}{c} {102}  & 
\multicolumn{1}{c} {R}  & 
\multicolumn{1}{c} {4.8}  & 
\multicolumn{1}{c} {1.1}  & 
\multicolumn{1}{c} {1.0}  & 
\multicolumn{1}{c} {-}  & 
\multicolumn{1}{c} {-}  & 
\multicolumn{1}{c} {-}  & 
\multicolumn{1}{c}{E/S0 **}  \\
\hline
\multicolumn{1}{c} {n2718b}  & 
\multicolumn{1}{c} {-17.8}  & 
\multicolumn{1}{c} {82}  & 
\multicolumn{1}{c} {R}  & 
\multicolumn{1}{c} {4.4}  & 
\multicolumn{1}{c} {1.0}  & 
\multicolumn{1}{c} {1.3}  & 
\multicolumn{1}{c} {-}  &
\multicolumn{1}{c} {-}  &
\multicolumn{1}{c} {-}  &
\multicolumn{1}{c}{E/S0 **}  \\
\hline
\multicolumn{1}{c} {n2775a}  & 
\multicolumn{1}{c} {-16.7}  & 
\multicolumn{1}{c} {401}  & 
\multicolumn{1}{c} {R}  & 
\multicolumn{1}{c} {-}  & 
\multicolumn{1}{c} {-}  & 
\multicolumn{1}{c} {-}  & 
\multicolumn{1}{c} {-}  & 
\multicolumn{1}{c} {-}  & 
\multicolumn{1}{c} {-}  & 
\multicolumn{1}{c}{Sc *}  \\
\hline
\multicolumn{1}{c} {n2775b}  & 
\multicolumn{1}{c} {-17.0}  & 
\multicolumn{1}{c} {416}  & 
\multicolumn{1}{c} {R}  & 
\multicolumn{1}{c} {-}  & 
\multicolumn{1}{c} {-}  & 
\multicolumn{1}{c} {-}  & 
\multicolumn{1}{c} {29.7}  & 
\multicolumn{1}{c} {2.5}  & 
\multicolumn{1}{c} {-}  & 
\multicolumn{1}{c}{Sc/Sd **}  \\
\hline
\multicolumn{1}{c} {n2775c}  & 
\multicolumn{1}{c} {-17.1}  & 
\multicolumn{1}{c} {64}  & 
\multicolumn{1}{c} {R}  & 
\multicolumn{1}{c} {-}  & 
\multicolumn{1}{c} {-}  & 
\multicolumn{1}{c} {-}  & 
\multicolumn{1}{c} {-}  & 
\multicolumn{1}{c} {-}  & 
\multicolumn{1}{c} {-}  & 
\multicolumn{1}{c}{S(B?)c **/Irr *}  \\
\hline
\multicolumn{1}{c} {a910a}  & 
\multicolumn{1}{c} {-16.0}  & 
\multicolumn{1}{c} {405}  & 
\multicolumn{1}{c} {R}  & 
\multicolumn{1}{c} {4.3}  & 
\multicolumn{1}{c} {1.0}  & 
\multicolumn{1}{c} {0.7}  & 
\multicolumn{1}{c} {-}  & 
\multicolumn{1}{c} {-}  & 
\multicolumn{1}{c} {-}  & 
\multicolumn{1}{c}{E/S0}  \\
\hline
\multicolumn{1}{c} {n2916a}  & 
\multicolumn{1}{c} {-17.8}  & 
\multicolumn{1}{c} {71}  & 
\multicolumn{1}{c} {R}  & 
\multicolumn{1}{c} {-}  & 
\multicolumn{1}{c} {-}  & 
\multicolumn{1}{c} {-}  & 
\multicolumn{1}{c} {-}  & 
\multicolumn{1}{c} {-}  & 
\multicolumn{1}{c} {-}  & 
\multicolumn{1}{c}{S (edge on)}  \\
\hline
\multicolumn{1}{c} {n2939a}  & 
\multicolumn{1}{c} {-15.7}  & 
\multicolumn{1}{c} {402}  & 
\multicolumn{1}{c} {R}  & 
\multicolumn{1}{c} {2.5}  & 
\multicolumn{1}{c} {0.5}  & 
\multicolumn{1}{c} {1.9}  & 
\multicolumn{1}{c} {-}  & 
\multicolumn{1}{c} {-}  & 
\multicolumn{1}{c} {-}  & 
\multicolumn{1}{c}{E/S0}  \\
\hline
\end{tabular}
\end{center}
\end{table}

\newpage
\newpage

\begin{table}
\begin{center}
\begin{tabular}{ccccccccccc}
\hline
\hline
\multicolumn{1}{c} {n3043a}  & 
\multicolumn{1}{c} {-16.5}  & 
\multicolumn{1}{c} {263}  & 
\multicolumn{1}{c} {R}  & 
\multicolumn{1}{c} {-}  & 
\multicolumn{1}{c} {-}  & 
\multicolumn{1}{c} {-}  & 
\multicolumn{1}{c} {4.5}  & 
\multicolumn{1}{c} {0.9}  & 
\multicolumn{1}{c} {-}  & 
\multicolumn{1}{c}{S (edge-on) *}  \\
\hline
\multicolumn{1}{c} {n3154a}  &
\multicolumn{1}{c} {-18.1}  &
\multicolumn{1}{c} {19}  &
\multicolumn{1}{c} {R}  &
\multicolumn{1}{c} {-}  &
\multicolumn{1}{c} {-}  &
\multicolumn{1}{c} {-}  &
\multicolumn{1}{c} {-}  &
\multicolumn{1}{c} {-}  & 
\multicolumn{1}{c} {-}  & 
\multicolumn{1}{c}{Sb/Sc}  \\
\hline
\multicolumn{1}{c} {n3629a}  & 
\multicolumn{1}{c} {-16.0}  & 
\multicolumn{1}{c} {409}  & 
\multicolumn{1}{c} {R}  & 
\multicolumn{1}{c} {2.8}  & 
\multicolumn{1}{c} {0.3}  & 
\multicolumn{1}{c} {0.6}  & 
\multicolumn{1}{c} {8.0}  & 
\multicolumn{1}{c} {0.8}  & 
\multicolumn{1}{c} {0.09}  & 
\multicolumn{1}{c}{Sa/Sb}  \\
\hline
\multicolumn{1}{c} {n3735a}  & 
\multicolumn{1}{c} {-17.6}  & 
\multicolumn{1}{c} {192}  & 
\multicolumn{1}{c} {R}  & 
\multicolumn{1}{c} {2.1}  & 
\multicolumn{1}{c} {0.4}  & 
\multicolumn{1}{c} {0.5}  & 
\multicolumn{1}{c} {41.4}  & 
\multicolumn{1}{c} {7.0}  & 
\multicolumn{1}{c} {0.04}  & 
\multicolumn{1}{c}{SBa/SBb}  \\
\hline
\multicolumn{1}{c} {n3735b}  & 
\multicolumn{1}{c} {-15.7}  & 
\multicolumn{1}{c} {119}  & 
\multicolumn{1}{c} {R}  & 
\multicolumn{1}{c} {3.6}  & 
\multicolumn{1}{c} {0.6}  & 
\multicolumn{1}{c} {0.7}  & 
\multicolumn{1}{c} {-}  & 
\multicolumn{1}{c} {-}  & 
\multicolumn{1}{c} {-}  & 
\multicolumn{1}{c}{E/S0}  \\
\hline
\multicolumn{1}{c} {n4030b}  & 
\multicolumn{1}{c} {-17.3}  & 
\multicolumn{1}{c} {414}  & 
\multicolumn{1}{c} {R}  & 
\multicolumn{1}{c} {1.4}  & 
\multicolumn{1}{c} {0.2}  & 
\multicolumn{1}{c} {6.7}  & 
\multicolumn{1}{c} {9.7}  & 
\multicolumn{1}{c} {1.2}  & 
\multicolumn{1}{c} {0.02}  & 
\multicolumn{1}{c}{Sc }  \\
\hline
\multicolumn{1}{c} {n4030c}  & 
\multicolumn{1}{c} {-17.3}  & 
\multicolumn{1}{c} {105}  & 
\multicolumn{1}{c} {R}  & 
\multicolumn{1}{c} {-}  & 
\multicolumn{1}{c} {-}  & 
\multicolumn{1}{c} {-}  & 
\multicolumn{1}{c} {-}  & 
\multicolumn{1}{c} {-}  & 
\multicolumn{1}{c} {-}  & 
\multicolumn{1}{c}{Irr (LSB)}  \\
\hline
\multicolumn{1}{c} {n4030d}  & 
\multicolumn{1}{c} {-17.3}  & 
\multicolumn{1}{c} {411}  & 
\multicolumn{1}{c} {R}  & 
\multicolumn{1}{c} {-}  & 
\multicolumn{1}{c} {-}  & 
\multicolumn{1}{c} {-}  & 
\multicolumn{1}{c} {-}  & 
\multicolumn{1}{c} {-}  & 
\multicolumn{1}{c} {-}  & 
\multicolumn{1}{c}{Sc **}  \\
\hline
\multicolumn{1}{c} {n4162a}  & 
\multicolumn{1}{c} {-17.3}  & 
\multicolumn{1}{c} {75}  & 
\multicolumn{1}{c} {R}  & 
\multicolumn{1}{c} {3.1}  & 
\multicolumn{1}{c} {0.5}  & 
\multicolumn{1}{c} {0.8}  & 
\multicolumn{1}{c} {6.3}  & 
\multicolumn{1}{c} {1.0}  & 
\multicolumn{1}{c} {0.6}  & 
\multicolumn{1}{c}{S0/Sa *}  \\
\hline
\multicolumn{1}{c} {n4541a}  & 
\multicolumn{1}{c} {-19.0}  & 
\multicolumn{1}{c} {193}  & 
\multicolumn{1}{c} {R}  & 
\multicolumn{1}{c} {-}  & 
\multicolumn{1}{c} {-}  & 
\multicolumn{1}{c} {-}  & 
\multicolumn{1}{c} {-}  & 
\multicolumn{1}{c} {-}  & 
\multicolumn{1}{c} {-}  & 
\multicolumn{1}{c}{S (edge on)}  \\
\hline
\multicolumn{1}{c} {n4541b}  & 
\multicolumn{1}{c} {-18.6}  & 
\multicolumn{1}{c} {228}  & 
\multicolumn{1}{c} {R}  & 
\multicolumn{1}{c} {2.1}  & 
\multicolumn{1}{c} {0.9}  & 
\multicolumn{1}{c} {5.2}  & 
\multicolumn{1}{c} {-}  & 
\multicolumn{1}{c} {-}  & 
\multicolumn{1}{c} {-}  & 
\multicolumn{1}{c}{E **}  \\
\hline
\multicolumn{1}{c} {n4541d}  & 
\multicolumn{1}{c} {-17.9}  & 
\multicolumn{1}{c} {95}  & 
\multicolumn{1}{c} {R}  & 
\multicolumn{1}{c} {3.5}  & 
\multicolumn{1}{c} {1.6}  & 
\multicolumn{1}{c} {0.8}  & 
\multicolumn{1}{c} {-}  & 
\multicolumn{1}{c} {-}  & 
\multicolumn{1}{c} {-}  & 
\multicolumn{1}{c}{SBc **}  \\
\hline
\multicolumn{1}{c} {n4541e}  & 
\multicolumn{1}{c} {-19.0}  & 
\multicolumn{1}{c} {217}  & 
\multicolumn{1}{c} {R}  & 
\multicolumn{1}{c} {3.1}  & 
\multicolumn{1}{c} {1.4}  & 
\multicolumn{1}{c} {0.8}  & 
\multicolumn{1}{c} {-}  & 
\multicolumn{1}{c} {-}  & 
\multicolumn{1}{c} {-}  & 
\multicolumn{1}{c}{S0/Sa **}  \\
\hline
\multicolumn{1}{c} {a1242a}  & 
\multicolumn{1}{c} {-18.3}  & 
\multicolumn{1}{c} {95}  & 
\multicolumn{1}{c} {R}  & 
\multicolumn{1}{c} {-}  & 
\multicolumn{1}{c} {-}  & 
\multicolumn{1}{c} {-}  & 
\multicolumn{1}{c} {2.9}  & 
\multicolumn{1}{c} {1.2}  & 
\multicolumn{1}{c} {-}  & 
\multicolumn{1}{c}{S (edge-on) *}  \\
\hline
\multicolumn{1}{c} {n4725a}  & 
\multicolumn{1}{c} {-18.2}  & 
\multicolumn{1}{c} {132}  & 
\multicolumn{1}{c} {I}  & 
\multicolumn{1}{c} {-}  & 
\multicolumn{1}{c} {-}  & 
\multicolumn{1}{c} {-}  & 
\multicolumn{1}{c} {-}  & 
\multicolumn{1}{c} {-}  & 
\multicolumn{1}{c} {-}  & 
\multicolumn{1}{c}{Sc **}  \\
\hline
\multicolumn{1}{c} {n4725b}  & 
\multicolumn{1}{c} {-15.8}  & 
\multicolumn{1}{c} {230}  & 
\multicolumn{1}{c} {I}  & 
\multicolumn{1}{c} {5.3}  & 
\multicolumn{1}{c} {0.4}  & 
\multicolumn{1}{c} {0.9}  & 
\multicolumn{1}{c} {-}  & 
\multicolumn{1}{c} {-}  & 
\multicolumn{1}{c} {-}  & 
\multicolumn{1}{c}{E}  \\
\hline
\multicolumn{1}{c} {n4939a}  & 
\multicolumn{1}{c} {-18.0}  & 
\multicolumn{1}{c} {415}  & 
\multicolumn{1}{c} {R}  & 
\multicolumn{1}{c} {5.2}  & 
\multicolumn{1}{c} {0.9}  & 
\multicolumn{1}{c} {0.7}  & 
\multicolumn{1}{c} {11.5}  & 
\multicolumn{1}{c} {2.0}  & 
\multicolumn{1}{c} {0.4}  & 
\multicolumn{1}{c}{Sc (barred?)}  \\
\hline
\multicolumn{1}{c} {n5248a}  & 
\multicolumn{1}{c} {-15.8}  & 
\multicolumn{1}{c} {150}  & 
\multicolumn{1}{c} {V}  & 
\multicolumn{1}{c} {-}  & 
\multicolumn{1}{c} {-}  & 
\multicolumn{1}{c} {-}  & 
\multicolumn{1}{c} {-}  & 
\multicolumn{1}{c} {-}  & 
\multicolumn{1}{c} {-}  & 
\multicolumn{1}{c}{Sd **/Irr (LSB)}  \\
\hline
\multicolumn{1}{c} {n5248b}  & 
\multicolumn{1}{c} {-15.8}  & 
\multicolumn{1}{c} {164}  & 
\multicolumn{1}{c} {I}  & 
\multicolumn{1}{c} {-}  & 
\multicolumn{1}{c} {-}  & 
\multicolumn{1}{c} {-}  & 
\multicolumn{1}{c} {7.4}  & 
\multicolumn{1}{c} {0.5}  & 
\multicolumn{1}{c} {-}  & 
\multicolumn{1}{c}{Sd **/Irr (LSB)}  \\
\hline
\multicolumn{1}{c} {n5899a}  & 
\multicolumn{1}{c} {-18.5}  & 
\multicolumn{1}{c} {106}  & 
\multicolumn{1}{c} {R}  & 
\multicolumn{1}{c} {2.4}  & 
\multicolumn{1}{c} {0.4}  & 
\multicolumn{1}{c} {0.5}  & 
\multicolumn{1}{c} {7.9}  & 
\multicolumn{1}{c} {1.3}  & 
\multicolumn{1}{c} {0.2}  & 
\multicolumn{1}{c}{Sb (dusty)}  \\
\hline
\multicolumn{1}{c} {n5921a}  & 
\multicolumn{1}{c} {-16.5}  & 
\multicolumn{1}{c} {239}  & 
\multicolumn{1}{c} {R}  & 
\multicolumn{1}{c} {-}  & 
\multicolumn{1}{c} {-}  & 
\multicolumn{1}{c} {-}  & 
\multicolumn{1}{c} {7.9}  & 
\multicolumn{1}{c} {0.9}  & 
\multicolumn{1}{c} {-}  & 
\multicolumn{1}{c}{S (edge-on)}  \\
\hline
\multicolumn{1}{c} {n5962a}  & 
\multicolumn{1}{c} {-15.3}  & 
\multicolumn{1}{c} {260}  & 
\multicolumn{1}{c} {I}  & 
\multicolumn{1}{c} {2.8}  & 
\multicolumn{1}{c} {0.3}  & 
\multicolumn{1}{c} {4.4}  & 
\multicolumn{1}{c} {-}  & 
\multicolumn{1}{c} {-}  & 
\multicolumn{1}{c} {-}  & 
\multicolumn{1}{c}{E/S0}  \\
\hline
\multicolumn{1}{c} {n5962d}  & 
\multicolumn{1}{c} {-17.5}  & 
\multicolumn{1}{c} {89}  & 
\multicolumn{1}{c} {R}  & 
\multicolumn{1}{c} {2.3}  & 
\multicolumn{1}{c} {0.3}  & 
\multicolumn{1}{c} {0.3}  & 
\multicolumn{1}{c} {9.9}  & 
\multicolumn{1}{c} {1.2}  & 
\multicolumn{1}{c} {0.03}  & 
\multicolumn{1}{c}{Sb/Sc}  \\
\hline
\multicolumn{1}{c} {n6181a}  & 
\multicolumn{1}{c} {-18.3}  & 
\multicolumn{1}{c} {257}  & 
\multicolumn{1}{c} {V}  & 
\multicolumn{1}{c} {-}  & 
\multicolumn{1}{c} {-}  & 
\multicolumn{1}{c} {-}  & 
\multicolumn{1}{c} {-}  & 
\multicolumn{1}{c} {-}  & 
\multicolumn{1}{c} {-}  & 
\multicolumn{1}{c}{Sc **}  \\
\hline
\multicolumn{1}{c} {n6384a}  & 
\multicolumn{1}{c} {-16.9}  & 
\multicolumn{1}{c} {474}  & 
\multicolumn{1}{c} {R}  & 
\multicolumn{1}{c} {7.1}  & 
\multicolumn{1}{c} {0.8}  & 
\multicolumn{1}{c} {0.6}  & 
\multicolumn{1}{c} {28.2}  & 
\multicolumn{1}{c} {3.1}  & 
\multicolumn{1}{c} {0.2}  & 
\multicolumn{1}{c}{SBb/SBc}  \\
\hline
\multicolumn{1}{c} {n7137a}  & 
\multicolumn{1}{c} {-15.0}  & 
\multicolumn{1}{c} {64}  & 
\multicolumn{1}{c} {R}  & 
\multicolumn{1}{c} {-}  & 
\multicolumn{1}{c} {-}  & 
\multicolumn{1}{c} {-}  & 
\multicolumn{1}{c} {-}  & 
\multicolumn{1}{c} {-}  & 
\multicolumn{1}{c} {-}  & 
\multicolumn{1}{c}{Irr (off-centre nuclei)}  \\
\hline
\multicolumn{1}{c} {n7177a}  & 
\multicolumn{1}{c} {-15.9}  & 
\multicolumn{1}{c} {151}  & 
\multicolumn{1}{c} {R}  & 
\multicolumn{1}{c} {-}  & 
\multicolumn{1}{c} {-}  & 
\multicolumn{1}{c} {-}  & 
\multicolumn{1}{c} {15.5}  & 
\multicolumn{1}{c} {1.1}  & 
\multicolumn{1}{c} {-}  & 
\multicolumn{1}{c}{Sb/Sc}  \\
\hline
\multicolumn{1}{c} {n7290a}  & 
\multicolumn{1}{c} {-15.9}  & 
\multicolumn{1}{c} {101}  & 
\multicolumn{1}{c} {R}  & 
\multicolumn{1}{c} {-}  & 
\multicolumn{1}{c} {-}  & 
\multicolumn{1}{c} {-}  & 
\multicolumn{1}{c} {-}  & 
\multicolumn{1}{c} {-}  & 
\multicolumn{1}{c} {-}  & 
\multicolumn{1}{c}{S **}  \\
\hline
\multicolumn{1}{c} {n7290b}  & 
\multicolumn{1}{c} {-15.9}  & 
\multicolumn{1}{c} {288}  & 
\multicolumn{1}{c} {R}  & 
\multicolumn{1}{c} {4.8}  & 
\multicolumn{1}{c} {0.9}  & 
\multicolumn{1}{c} {0.8}  & 
\multicolumn{1}{c} {-}  & 
\multicolumn{1}{c} {-}  & 
\multicolumn{1}{c} {-}  & 
\multicolumn{1}{c}{E (nucleated)}  \\
\hline
\multicolumn{1}{c} {n7678a}  & 
\multicolumn{1}{c} {-18.0}  & 
\multicolumn{1}{c} {165}  & 
\multicolumn{1}{c} {R}  & 
\multicolumn{1}{c} {2.4}  & 
\multicolumn{1}{c} {0.6}  & 
\multicolumn{1}{c} {0.6}  & 
\multicolumn{1}{c} {9.4}  & 
\multicolumn{1}{c} {2.4}  & 
\multicolumn{1}{c} {0.1}  & 
\multicolumn{1}{c}{Sb/Sc}  \\
\hline
\end{tabular}
\end{center}
\end{table}


\newpage

\begin{thebibliography}{}
\bibitem[]{}Aguerri, J. A. L., Balcells, M., \& Peletier, R. F. 2001, A\&A, 367, 428
\bibitem[]{}Barnes, J. E., \& Hernquist, L. E. 1991, ApJ, 370, L65
\bibitem[]{}Benson, A. J., Frenk, C. S., Lacey, C. G., Baugh, C. M., \& Cole, S. 2001, astro-ph/0108218
\bibitem[]{}Blitz, L., Spergel, D., Teuben, P., Hartmann, D., \& Burton, W. B.  1999, ApJ, 574, 818
\bibitem[]{}Bode, P., J. P. Ostriker, \& Turok, N. 2001, ApJ, 556, 93
\bibitem[]{}Bullock, J. S., Kravtsov, A. V., Weinberg, D. H. 2000, ApJ, 539, 517
\bibitem[]{}Braun, R., \&  Burton, W. B. 1999, AA, 341, 437
\bibitem[]{}Capaccioli, M. 1989, in The World of Galaxies, ed.  H. G. Corwin \& L. Bottinelli (Berlin: Springer-Verlag), p. 208
\bibitem[]{}Carignan, C., Cote, S., Freeman, K. C., \& Quinn, P. J. 1997, AJ, 113, 1585
\bibitem[]{}Cote, S., Freeman, K. C., Carignan, C., \& Quinn, P. J. 1997, AJ, 114, 1313
\bibitem[]{}Frenk, C. S., White, S. D. M., Davis, M., \& Efstathiou, G. 1988, ApJ, 327, 507
\bibitem[]{}Gorham, P. W., van Zee, L., Unwin, S. C., \& Jacobs, C. 2000, AJ, 119, 1677
\bibitem[]{}Grebel, E. K. 2001, astro-ph/0011048 
\bibitem[]{}Holmberg, E. 1969, Ark.  Astron., 5, 305
\bibitem[]{}Kauffmann, G., White, S. D. M., \& Guiderdoni, B. 1993, MNRAS, 264, 201
\bibitem[]{}Klypin, A., Kravtsov, A. V., Valenzuela, O., \& Prada, F. 1999, ApJ, 522, 82
\bibitem[]{}Mateo, M. L. 1998, ARA\&A, 36, 435
\bibitem[]{}Mihos, J. C., \& Hernquist, L. 1994, ApJ, 425, L13
\bibitem[]{}Moore, B., Ghigna, S., Governato, F., Lake, G., Quinn, T., Stadel, J., \& Tozzi, P. 1999, ApJ, 524, L19
\bibitem[]{}Page, T. 1952, 116, 63
\bibitem[]{}Pritchet, C. J., \& van den Bergh, S. 1999, AJ, 118, 883
\bibitem[S\'ersic(1968)]{ser68}  S\'ersic, J.  1968, Atlas de Galaxias Australes C\'ordoba: Obs. Astron\'omico
\bibitem[]{}Quinn, P. J., \& Goodman, J. (1986), ApJ, 309, 472
\bibitem[]{}Somerville, R. S. 2001, astro-ph/0107507
\bibitem[]{}Thakar, A. R., Ryden, B. S. 1996, ApJ 461, 55
\bibitem[]{}Toth, G., \& Ostriker, J. P. 1992, ApJ, 389, 5
\bibitem[]{}Trujillo, I., Aguerri, J. A. L., Cepa, J., \& Guti\'errez, C. M. 2001, MNRAS, 321, 269
\bibitem[]{}van den Bergh, S. 2000, PASP, 112, 529
\bibitem[]{}Velazquez, H., \& White, S. D. M. 1999, MNRAS, 304, 254
\bibitem[]{}Walker, I. R., Mihos, J. C., \& Hernquist, L. 1996, ApJ, 460, 121
\bibitem[]{}Weinberg, M. D. 1997, ApJ, 478, 435
\bibitem[]{}Zaritsky, D. 1995, ApJ, 448, L17
\bibitem[]{}Zaritsky, D., Smith, R., Frenk, C., \& White, S. D. M. 1993, ApJ, 405, 464
\bibitem[]{}Zaritsky, D., Smith, R., Frenk, C., \& White, S. D. M. 1997, ApJ, 478, 39 
\end{thebibliography}
\end{document}